\documentclass[aîáð,aip,twocolumn,reprint,preprintnumbers,amsmath,amssymb]{revtex4-1}

\usepackage{graphicx}        
\usepackage{longtable}       
\usepackage{amsfonts}        
\usepackage{amsmath}         
\usepackage{amssymb}         
\usepackage{bm}              
\usepackage{color}           
\usepackage{dcolumn}         
\usepackage{epsfig}          
\usepackage{hyperref}        

\newcommand{\beq}{\begin{equation}}
\newcommand{\eeq}{\end{equation}}
\newcommand{\bea}{\begin{eqnarray}}
\newcommand{\eea}{\end{eqnarray}}

\newcommand{\APL}{Appl. Phys. Lett. }

\newcommand{\JAP}{J. Appl. Phys. }

\newcommand{\PRB}{Phys. Rev. B }

\newcommand{\PRL}{Phys. Rev. Lett. }

\begin{document}

\title{Design of $L2_1$-type antiferromagnetic semiconducting full-Heusler compounds:
A first principles DFT+$GW$ study}

\author{M. Tas$^1$}\email{tasm236@gmail.com}

\author{E. \c{S}a\c{s}{\i}o\u{g}lu$^2$}
\altaffiliation{Present address: Institut f\"ur Physik,
Martin-Luther-Universit\"at Halle-Wittenberg, D-06099 Halle
(Saale) Germany}

\author{C. Friedrich$^2$}

\author{S. Bl\"{u}gel$^2$}

\author{I. Galanakis$^1$}\email{galanakis@upatras.gr}

\affiliation{$^1$Department of Materials Science, School of
Natural Sciences, University of Patras, GR-26504 Patra, Greece\\
$^2$Peter Gr\"{u}nberg Institut and Institute for Advanced
Simulation, Forschungszentrum J\"{u}lich and JARA, 52425
J\"{u}lich, Germany}

\begin{abstract}
Antiferromagnetic spintronics is an on-going growing field of research.
Employing both standard density functional theory and the $GW$ approximation
within the framework of the FLAPW method, we study the electronic and
magnetic properties of seven potential antiferromagnetic semiconducting
Heusler compounds with 18 (or 28 when Zn is present) valence electrons per
unit cell. We show that in these compounds G-type antiferromagnetism is the
ground state and that they are all either semiconductors (Cr$_2$ScP,
Cr$_2$TiZn, V$_2$ScP, V$_2$TiSi, and V$_3$Al) or semimetals (Mn$_2$MgZn and
Mn$_2$NaAl). The many-body corrections have a minimal effect on the
electronic band structure with respect to the standard electronic structure
calculations.
\end{abstract}

\maketitle

\section{Introduction}

Among the subfields of spintronics, the so-called
``antiferromagnetic spintronics" is constantly growing during the
recent years.\cite{Review1,Review2} This new research field deals
with the implementation of antiferromagnetic semiconductors in
conventional spintronic devices either as a substitute to the
ferromagnetic materials or in heterostructures with the latter,
\textit{e.g.} in spin torque-transfer (STT) magnetic
memories.\cite{Review2} Antiferromagnetic semiconductors have also
been proposed in order to control the properties of magnetic
topological insulators unveiling new avenues towards
dissipationless topological antiferromagnetic
spintronics.\cite{He2016} Antiferromagnets, contrary to
ferromagnets, create vanishing stray fields leading to minimal
energy losses. On the other hand the current-control of the
magnetic information stored in antiferromagnetic materials still
needs to be studied in depth.\cite{Review1,Review2} Main
challenges in the antiferromagnetic spintronics involve their
coherent growth on top of ferromagnets, like in the STT memories,
and the discovery of antiferromagnets with high N\'eel temperature
($T_{\mathrm{N}}$) is requested for realistic devices since most
have $T_{\mathrm{N}}$ far below the room
temperature.\cite{Review2}

Heusler compounds are playing a central role in the development of
spintronics and magnetoelectronics.\cite{Perspectives,book} These
ternary and quaternary intermetallic compounds present a large
variety of magnetic behaviors and their magnetic properties are
implicitly connected to their electronic
properties.\cite{Galanakis1,Galanakis2,Galanakis3,Galanakis4}
Although several studies have been devoted to Heusler compounds
(for a review see Refs. \onlinecite{Felser1,Felser2}), the ongoing
research reveals new properties with potential interest for
applications.\cite{Perspectives} Among them there are
\textit{ab-initio} studies which have identified few Heusler
compounds exhibiting zero net magnetization. The latter are   made
of magnetic constituents and are usually defined as half-metallic
fully compensated ferrimagnets (also known as half-metallic
antiferromagnets),\cite{Wurmehl2006,Tirpanci2013} and Cr$_2$CoGa,
which belongs to this class of materials, has been successfully
grown.\cite{Cr2CoGa,Cr2CoGab} But, these materials are metals and
at finite temperature the atomic spin magnetic moments are not
compensated anymore leading to usual ferrimagnetic
behavior.\cite{Wurmehl2006} Thus, such materials do not fulfil the
requirements for antiferromagnetic spintronics.

\begin{table*}
\caption{PBE calculated lattice constants, $a_\textrm{eq}$ in \AA\
(we published the value for V$_3$Al in Ref.
\onlinecite{Galanakis2016}), and difference of the total energies
(in eV) between the antiferromagnetic (AFM) and non-magnetic (NM)
states, $\Delta \textrm{E}^{\textrm{AFM-NM}}$, as well as the
difference between the AFM and ferromagnetic (FM) configurations,
$\Delta \textrm{E}^{\textrm{AFM-FM}}$. The AFM state is the ground
state in all cases. The other columns present the PBE calculated
spin magnetic moments in $\mu_{\textrm{B}}$ for the Heusler
compounds under study in the AFM case. We also present in
parenthesis the results for the FM coupling of the spin magnetic
moments. We use the symbols $A$ and $C$ to denote the two
transition metal atoms sitting at different sites.  For V$_2$TiSi
and V$_3$Al we could not get a FM solution irrespective of the
starting distribution of the atomic spin magnetic moments. Note
that both the energy differences and the total spin magnetic
moments are given per formula unit.}

\label{table1}
\begin{ruledtabular}
\begin{tabular}{lcccccccc}
X$_2$YZ & $a_{\textrm{eq}}$(\AA) & $\Delta \textrm{E}^{\textrm{AFM-NM}}$ &
$\Delta \textrm{E}^{\textrm{AFM-FM}}$ & $m^{\textrm{X(A)}}$ & $m^{\textrm{X(C)}}$ &
$m^{\textrm{Y}}$ & $m^{\textrm{Z}}$ & $m^{\textrm{total}}$ \\
\hline
Cr$_2$ScAl & 6.39 & $-1.946$ & $-0.756$ & 3.306 (2.677) & $-3.306$ (2.677) & 0 ~ (0.255)  & 0 ($-0.086$) & 0 (5.522) \\
Cr$_2$TiZn & 6.14 & $-1.162$ & $-0.829$ & 2.918 (2.380) & $-2.918$ (2.380) & 0 ($-0.018$) & 0 ~ (0.019)  & 0 (4.761) \\
Mn$_2$NaAl & 6.45 & $-2.831$ & $-0.162$ & 3.867 (3.671) & $-3.867$ (3.671) & 0 ($-0.006$) & 0 ($-0.217$) & 0 (7.119) \\
Mn$_2$MgZn & 6.23 & $-2.310$ & $-0.130$ & 3.582 (3.296) & $-3.582$ (3.296) & 0 ($-0.061$) & 0 ($-0.115$) & 0 (6.415) \\
V$_2$ScP   & 6.17 & $-0.692$ & $-0.552$ & 2.155 (1.405) & $-2.155$ (1.405) & 0 ~ (0.488)  & 0 ($-0.026$) & 0 (3.272) \\
V$_2$TiSi  & 6.10 & $-0.379$ & ---      & 1.776         & $-1.776$         & 0            & 0            & 0         \\
V$_3$Al    & 6.09 & $-0.131$ & ---      & 1.384         & $-1.384$         & 0            & 0            & 0         \\
\end{tabular}
\end{ruledtabular}
\end{table*}

The large number of possible combinations of chemical elements in
Heusler compounds leads unavoidably to the conclusion that it
could be feasible to identify also antiferromagnetic
semiconductors among them. Such materials would combine the large
critical temperature usually exhibited by Heusler compounds to the
coherent growth on top of other ferromagnetic Heusler compounds
since most of them crystallize in the same cubic structure.
Motivated by the experimental detection of the antiferromagnetic
semiconducting behavior in V$_3$Al, a Heusler compound with 18
valence electron crystallizing in the $D0_3$ lattice structure, we
search for other potential antiferromagnetic semiconducting
Heusler compounds with 18 valence electrons per unit cell. We
identified six candidates Cr$_2$ScAl, Cr$_2$TiZn, Mn$_2$NaAl,
Mn$_2$MgZn, V$_2$ScP, and V$_2$TiSi.  Cr$_2$TiZn and Mn$_2$MgZn
have actually 28 valence electrons due to the presence of the Zn
atom. But Zn's 3\textit{d} orbitals  are completely occupied
creating isolated bands low in energy and, thus, can be considered
semicore states not affecting the studied properties. Therefore,
we will refer to them as 18-valence electrons compounds, as well.
We should note finally that, as it is well-known for half-metallic
or magnetic semiconducting Heusler compounds, the substitution of
an element by an isovalent one (\textit{i.e.} same number of
valence electrons) does not alter the electronic and the magnetic
properties of the compounds since the latter depend on the total
number of valence electrons in the unit
cell.\cite{Galanakis2,Galanakis3,Galanakis4} Thus, \textit{e.g}
Mn$_2$LiAl and Mn$_2$KAl show similar properties to Mn$_2$NaAl.

\section{Computational method}

We have performed simulations of the electronic and magnetic
properties of the six compounds and V$_3$Al using the
density-functional-theory (DFT) based on the full-potential
linearized augmented-plane-wave (FLAPW) method as implemented in
the \texttt{FLEUR} code \cite{Fleur} within the generalized
gradient approximation (GGA) of the exchange-correlation potential
as parameterized by Perdew, Burke and Ernzerhof (PBE).\cite{PBE}
Since such calculations are restricted to the ground-state
properties and often fail in describing the band gap of
semiconductors, we have also employed the $GW$ approximation using
the PBE results as input to perform one-shot $GW$ calculations
using the Spex code.\cite{SPEX} Details of the calculations are
identical to the ones in Ref. \onlinecite{Tas}, where non-magnetic
semiconducting Heusler compounds were studied. Also, since our
calculations refer to the zero-temperature ground-state, the
description of the magnetic properties at finite temperatures
could be more complicated; \textit{e.g} for V$_3$Al the
longitudinal spin-fluctuations play a decisive role in the
formation of the vanadium magnetic moments.\cite{Khmelevskyi}

Prior to presenting our results in the next section, we should
also discuss the possibility of growing experimentally the
compounds under study. V$_3$Al is the only one which has been
grown experimentally in the cubic Heusler structure as a
film\cite{Jamer} confirming the predictions of Skaftouros et al.
in Ref. \onlinecite{SkaftourosAPL}; note that the stable lattice
structure of V$_3$Al is the A15 and V$3$Al in this structure is a
well-known superconductor.\cite{A15}  Searching the Open Quantum
Materials Database we found that all other compounds under study,
with the exception of V$_2$ScP, can exist experimentally although
the cubic $L2_1$ lattice structure is not the most stable phase;
in the case of the V-Sc-P phase diagram no ternary compound was
suggested to exist.\cite{oqmd1,oqmd2} Thus we expect that modern
growth instruments and techniques would allow the successful
growth of the compounds under study in the present article as was
recently the case of the quaternary antiferromagnetic
semiconducting Heusler (CrV)TiAl. For the latter, its existence
was suggested using ab-initio calculations in Ref.
\onlinecite{CrVTiAl} and recently Stephen and collaborators have
grown it in the form a film confirming also its theoretical
predicted antiferromagnetic semiconducting
character.\cite{Stephen}

\begin{figure}
\begin{center}
\includegraphics[width=\columnwidth]{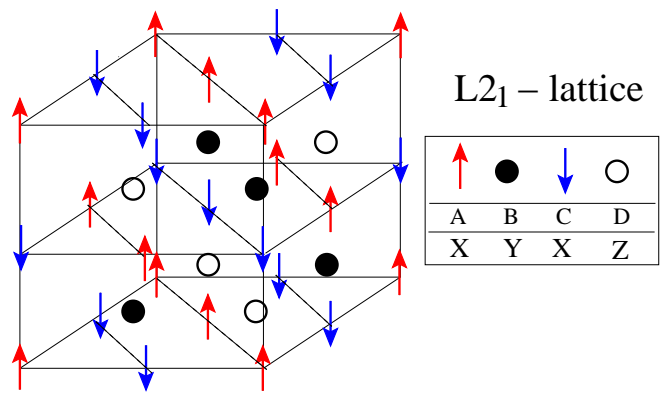}
\end{center}
\vspace{-0.6cm} \caption{(Color online) Schematic representation
of the $L2_1$ lattice structure. For the two inequivalent X atoms
we also demonstrate the direction of the spin magnetic moments in
the antiferromagnetic configuration.} \label{fig1}
\end{figure}

\section{Results and discussion}

 We begin our discussion by presenting the
ground state properties as obtained by the PBE calculations. The
compounds under study have the formula X$_2$YZ and crystallize in
the $L2_1$ cubic lattice structure which is presented in
Fig.~\ref{fig1}. It is actually a fcc lattice with four atoms as
basis along the diagonal: X atoms at the A and C sites at  (0 0 0)
and ($\frac{1}{2}$ $\frac{1}{2}$ $\frac{1}{2}$), Y atoms at the B
site at ($\frac{1}{4}$ $\frac{1}{4}$ $\frac{1}{4}$), and Z atoms
at the D ($\frac{3}{4}$ $\frac{3}{4}$ $\frac{3}{4}$) site.  The
exchange constants of the 3\textit{d}-transition metals obey the
semi-phenomenological Slater-Bethe-N\'eel curve which predicts
antiferromagnetism for the early transition metals with less than
half-filled shells, like V, Cr and Mn, for small interatomic
distances.\cite{Skomski,Jiles} In the compounds under study the X
atoms at the A and C sites are second neighbors and their distance
is exactly half the lattice constant as can be seen in Fig.
\ref{fig1}. This value is for all compounds under study slightly
larger than 3 \AA , as we can deduce from the lattice constants
presented in Table \ref{table1}, and is not far away from the 2.52
\AA\ which is the distance between the Cr atoms in the
antiferromagnetic bcc Cr. Thus we expect for the compounds under
study that the X atoms at the A and C sites show an
antiferromagnetic coupling of their spin magnetic moments.
Assuming that the ground state is the G-type antiferromagnetic
state shown in Fig.~\ref{fig1},\cite{Galanakis2016} we first
determined the equilibrium lattice constants using total energy
calculations. The obtained results are presented in the first
column of Table~\ref{table1}. Lattice constants in all cases
exceed 6 \AA , nearly reaching 6.5 \AA\ for Mn$_2$MgZn. We then
established the stability of the antiferromagnetic state by
performing calculations where we imposed our systems to be either
non-magnetic or ferromagnetic. We note that we could not get
ferromagnetic ground-states for V$_2$TiSi and V$_3$Al.

In the second column of Table~\ref{table1} we present the
difference in total energy between the antiferromagnetic and the
non-magnetic states. The minus sign implies that the
antiferromagnetic state is more favorable. V$_3$Al shows the
smallest energy difference of 0.131 eV. For the Cr- and Mn-based
compounds, the energy difference is one order of magnitude larger,
so the magnetic state is considerably more stable. In the third
column we present the difference in total energy between the
antiferromagnetic and the ferromagnetic configurations calculated
at the equilibrium lattice constant. For the Cr-based compounds,
the energy difference shows a pretty large value of about $-0.8$
eV in favor of the antiferromagnetic configuration, while in the
Mn-based compounds it is much smaller $-0.13$ and $-0.16$ eV. The
value of $-0.13$ eV is almost identical to V$_3$Al for which the
N\'eel temperature ($T_{\mathrm{N}}$) has been calculated to be
988 K,\cite{Galanakis2016} and thus for the rest of the compounds
we expect significantly larger $T_{\mathrm{N}}$ values making the
antiferromagnetism a very robust property of the compounds under
study even at elevated temperatures.

\begin{table*}
\caption{Calculated  PBE (in parenthesis) and \textit{GW} energy
band gaps and transition energies (all in eV) between certain
high-symmetry points for  the antiferromagnetic semiconducting Cr-
and V-based studied materials. \label{table2}}
\begin{ruledtabular}
\begin{tabular}{lccccc}
Compound  & E$_{\textrm{g}}^{GW}$ (E$_{\textrm{g}}^\mathrm{PBE}$)
& $\Gamma \rightarrow \Gamma$ &
X$\rightarrow$X & $\Gamma \rightarrow$X & X$\rightarrow \Gamma$ \\
\hline

Cr$_2$ScAl  &  0.893 (0.674)  & 1.242 (1.143) & 2.212 (1.676) & 1.032 (0.882) & 2.422 (1.938)  \\
Cr$_2$TiZn  &  0.800 (0.586)  & 1.782 (1.681) & 1.417 (1.028) & 1.062 (1.079) & 2.137 (1.630)  \\
V$_2$ScP    &  0.034 (0.123)  & 0.034 (0.123) & 1.995 (2.124) & 1.383 (1.350) & 0.646 (0.897)  \\
V$_2$TiSi   &  0.427 (0.383)  & 1.935 (1.984) & 1.736 (1.871) & 1.344 (1.350) & 2.328 (2.506)  \\
V$_3$Al     &  0.065 (0.081)  & 2.159 (2.320) & 1.575 (1.185) & 1.171 (1.079)& 2.563 (2.426)  \\

\end{tabular}
\end{ruledtabular}
\end{table*}

In Table \ref{table1} we present the atomic spin magnetic moments
in $\mu_{\mathrm{B}}$ for both the antiferromagnetic and the
ferromagnetic configurations; the latter in parenthesis. In the
G-type antiferromagnetism case presented in Fig.~\ref{fig1},
consecutive (111) planes of X atoms have antiparallel moments, and
the Y and Z atoms exhibit a zero spin magnetic moment due to
symmetry reasons, \textit{i.e.} they are surrounded by four X
atoms sitting at A sites with positive and four X atoms at C sites
with negative spin magnetic moments. The absolute values of the
atomic spin magnetic moments of the X atoms are pretty high
exceeding even 3.5 $\mu_{\mathrm{B}}$ in the case of the Mn atoms.
These values are significantly larger than the corresponding spin
magnetic moments in the ferromagnetic case. The total spin
magnetic moment for all antiferromagnetic compounds is exactly
zero. This is compatible with the appearance of semiconducting
behavior although it does not guarantee it as we will discuss
later on.

\begin{figure}
\begin{center}
\includegraphics[width=\columnwidth]{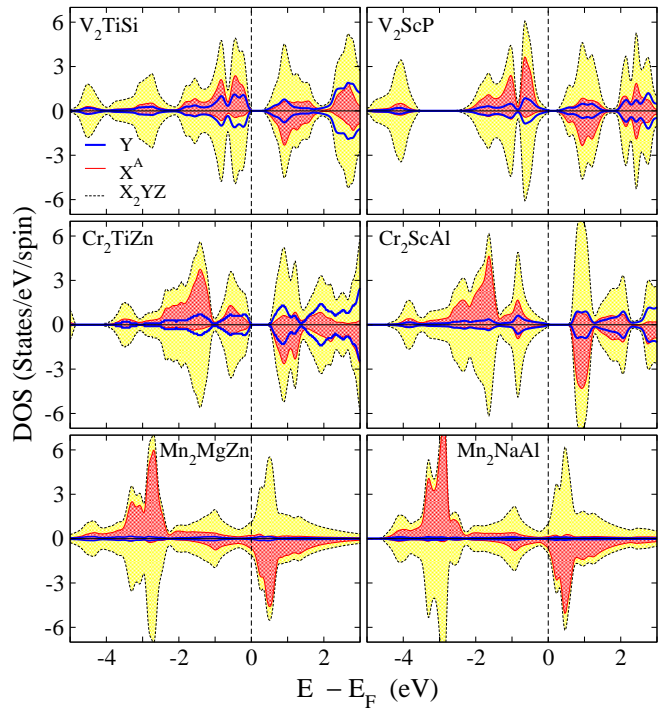}
\end{center}
\vspace{-0.6cm} \caption{(Color online) Atom-resolved and total DOS for the
X$_2$YZ compounds under study. The zero energy value corresponds to the Fermi
level. Positive (negative) DOS values correspond to the spin-up (spin-down)
electrons.}
\label{fig2}
\end{figure}

We  briefly discuss the atom-resolved electronic properties. In
Fig.~\ref{fig2} we present the density of states (DOS) for all
studied compounds. In the case of the V- and Cr-based compounds
the orbitals of the Y atoms have a significant weight at the same
region as the \textit{d}-orbitals of the X atoms (V or Cr), and
thus hybridization is pretty strong. On the contrary, in the case
of the two Mn-based compounds the situation is clearly different.
Occupied Mn \textit{d}-orbitals are almost of a unique spin
character due to the larger exchange splitting, and thus they are
located much lower in energy leading to a localization of the spin
magnetic moment similar to other Mn-based Heusler
compounds.\cite{Sommers} Simultaneously the Y atoms (Mg or Na)
have vanishing weight at the energy region of the Mn
\textit{d}-orbitals. Thus hybridization is very weak in the
Mn$_2$MgZn and Mn$_2$NaAl compounds.

\begin{figure}
\begin{center}
\includegraphics[width=\columnwidth]{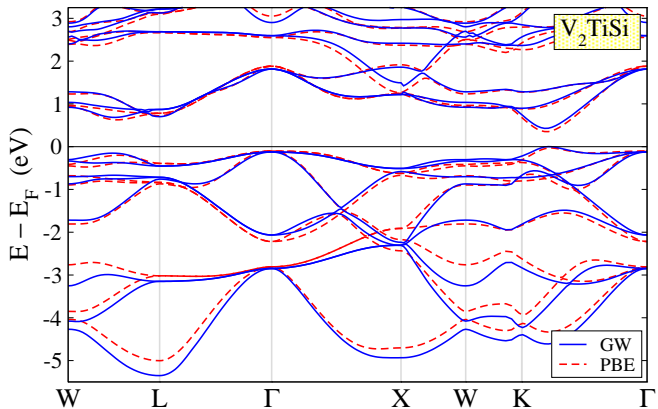}
\end{center}
\vspace{-0.6cm} \caption{(Color online) Calculated electronic band
structure of V$_2$TiSi along the high-symmetry directions in the
first Brillouin zone using either the PBE (red dashed line) or the
$GW$ (blue solid line) approximations. The zero energy value
denotes the Fermi level.} \label{fig3}
\end{figure}

Next we focus on the calculation of the electronic band structure
of the compounds under study as well as the determination of the
energy band gaps and of the transition energies. For the $GW$
calculations we have used a $8 \times 8 \times 8$ grid to carry
out the calculations which is dense enough to accurately reproduce
the band structure as shown also in Ref.~\onlinecite{Tas}. We
present  all obtained energy values within both the PBE and the
$GW$ approximations in Table~\ref{table2}. Moreover in
Fig.~\ref{fig3} we present, as an example, the band structure for
V$_2$TiSi along the high symmetry lines in the Brillouin zone
using both PBE and $GW$. The band structures of the other
compounds look similar.

Prior to the discussion of the results, we note that all compounds
studied have 18 valence electrons, 9 per spin direction; thus the
hybridizations scheme is similar to that of V$_3$A in
Ref.~\onlinecite{Galanakis2016}. In the case of semiconducting
behavior, the gap is created between the occupied
triple-degenerate $t_{2g}$ orbitals obeying both the octahedral
and tetrahedral symmetry groups and being extended over all X and
Y sites, and the triple-degenerate $t_{1u}$ orbitals obeying
exclusively the octahedral symmetry and thus being located only at
the X atoms.

We start our presentation of the electronic properties with the
three V based compounds. All three compounds are G-type
antiferromagnetic semiconductors as can be deduced from their
total DOS shown in Fig.~\ref{fig2}. The band structure of
V$_2$TiSi is presented in Fig.~\ref{fig3} where we observe that
overall PBE and $GW$ present a qualitatively similar band
structure picture especially around the Fermi level, as it was the
case for the non-magnetic semiconductors in Ref.~\onlinecite{Tas}.
The use of $GW$ provides a direct band gap close to the K point
which is slightly larger than the PBE calculated band gap, 0.427
eV with respect to 0.383 eV. Hence, V$_2$TiSi can be classified as
a narrow-band antiferromagnetic semiconductor. Although the energy
gap is very small, the transition energies presented in
Table~\ref{table2} are very large, and away from the point in
\textbf{k}-space, where the direct gap exists, large photon
energies are needed to excite electrons since the latter in the
elastic regime conserve their \textbf{k}-value. The $GW$
transition energies are smaller than the PBE ones contrary to the
usual effect of quasiparticles on the band structure of
semiconducting materials. This is also observed in other
semiconductors like some non-magnetic semiconducting Heusler
compounds.\cite{Tas}

V$_2$ScP, contrary to V$_2$TiSi, exhibits a very narrow direct gap
of 0.034 eV at the $\Gamma$ point, while the transition energy at
the X point is quite large (about 2 eV). The indirect $\Gamma
\rightarrow$X and X$\rightarrow \Gamma$ gaps are two thirds and
one third of the X$\rightarrow$X transition energy, respectively.
Therefore, V$_2$ScP can be classified as an almost-gapless
semiconductor possessing a direct gap exactly at the $\Gamma$
point. V$_3$Al within $GW$ is found to be an almost-gapless
semiconductor in agreement with previous
calculations.\cite{Galanakis2016} Although its total DOS (note
shown here) resembles that of V$_2$ScP, its band structure
actually resembles that of V$_2$TiSi with a small direct gap
between the K and $\Gamma$ points. The width of the gap is 0.081
eV within PBE and slightly smaller (0.065 eV) within the $GW$.
Transition energies for V$_3$Al are significantly larger than the
band gap as was the case for V$_2$TiSi.

Cr$_2$ScAl and Co$_2$TiZn also show a semiconducting behavior.
Their band structure resembles that of V$_2$TiSi in
Fig.~\ref{fig3}, but with a much flatter valence band near the
$\Gamma$ point. For both compounds we have an indirect gap with
the minimum of the conduction band being located between the K and
$\Gamma$ points. The maximum of the valence band is exactly at the
$\Gamma$ point for Cr$_2$ScAl and between $\Gamma$ and X for
Cr$_2$TiZn. The band gap for both compounds is 0.8-0.9 eV. Thus
they should be classified as antiferromagnetic narrow-band
semiconductors. The $GW$ self-energy correction gives larger band
gaps and transition energies than PBE, contrary to what we have
found for the V compounds, while being in accordance with the
usual self-energy effect on the band structure of semiconductors.

The two Mn compounds show a different behavior. As obvious from
Fig.~\ref{fig2}, they are semimetals with a region of low DOS
around the Fermi level. The Fermi level crosses both the
conduction and valence bands creating electron and hole pockets.
The bottom of the conduction band is 0.143 and 0.127 eV below the
Fermi level for Mn$_2$NaAl and Mn$_2$MgZn, respectively (the PBE
values are 0.174 eV and 0.106 eV). The transition energy at the
$\Gamma$ point is 0.628 eV and 0.548 eV for the two compounds,
respectively (0.594 eV and 0.548 eV using PBE).

Finally, we have also calculated the Cr$_2$ScAl and Cr$_2$TiZn
compounds assuming the so-called inverse (or $XA$ or $Xa$) lattice
structure where the sequence of the atoms changes with respect to
the usual $L2_1$ lattice structure; the sequence along the
diagonal of the cube in Fig. \ref{fig1} is now
Cr-Cr-Sc(Ti)-Al(Zn). The inverse structure occurs usually when the
valence of the Y atom in the X$_2$YZ compound is larger than the
valence of the X atoms. We found that the inverse structure is
higher in energy with respect to the $L2_1$ by 0.1-0.2 eV and thus
the $L2_1$ is energetically favorable. The Cr atoms, which are now
nearest neighbors, possess large antiparallel spin magnetic
moments slightly smaller than in the $L2_1$ case, but now also the
Sc(Ti) atoms possess considerable spin magnetic moments which are
parallel to the spin moments of the Cr atoms at the B sites and
antiparallel to the spin moments of the Cr atoms at the A sites,
similar to the situation occurring in other full-Heusler
crystallizing in the inverse structure like
Mn$_2$CoAl.\cite{SkaftourosAPL} The total spin magnetic moment
slightly deviates from the zero value and the compound is now a
usual almost fully-compensated ferrimagnetic half-metal with a low
DOS in the spin-up band structure and a gap in the spin-down band
structures instead of an antiferromagnetic semiconductor as in the
$L2_1$ structure.

\section{Summary and conclusions}

In conclusion, we have studied the quasi-particle band structure
of several full-Heusler compounds with 18 (or 28 if Zn is present)
valence electrons susceptible of being antiferromagnetic
semiconductors. In all cases the favorable magnetic configuration
is the G-type antiferromagnetic one with large spin magnetic
moments at the transition metal sites. Our results suggest that
the V-based compounds (V$_3$Al, V$_2$ScP, V$_2$TiSi) and the
Cr-based ones (Cr$_2$ScAl, Cr$_2$TiZn) are actually almost-gapless
or narrow-gap semiconductors, while the Mn-based compounds
(Mn$_2$NaAl, Mn$_2$MgZn) are semimetals. The $GW$ approximation
leads to band gaps and transition energies that do not deviate
much from the PBE ones. Thus the standard DFT based
first-principles calculations are reliable for describing the
electronic and magnetic properties of these materials. We expect
our results to further intensify the interest in antiferromagnetic
spintronics and the use of Heusler compounds in this new research
field.

\begin{acknowledgments}
Authors wish to thank \c{S}. T{\i}rpanc{\i} for fruitful
discussions.
\end{acknowledgments}

\end{document}